\documentstyle[preprint,pra,aps]{revtex}
\begin{document}
\draft
\title{Cherenkov radiation of superluminal particles}
\author{Daniel Rohrlich$^{1}$\thanks{E-mail:  \tt
rohrlich@wicc.weizmann.ac.il} and Yakir Aharonov$^2$}
\address{$^1$Department of Condensed Matter Physics, Weizmann Institute
of Science, Rehovot 76100 Israel}
\address{$^2$School of Physics and Astronomy, Tel Aviv University,
Ramat Aviv 69978 Tel Aviv, Israel and Department of Physics, University
of South Carolina, Columbia, South Carolina 29208}
\date{printed \today}
\maketitle
\begin{abstract}
Any charged particle moving faster than light through a medium emits
Cherenkov radiation.  We show that charged particles moving faster
than light through the {\it vacuum} emit Cherenkov radiation.  How can
a particle move faster than light?  The {\it weak} speed of a charged
particle can exceed the speed of light.  By definition, the weak
velocity $\langle {\bf v}\rangle_w$ is $\langle \Psi_{fin} \vert {\bf
v} \vert \Psi_{in} \rangle/ \langle \Psi_{fin} \vert \Psi_{in} \rangle$,
where ${\bf v}$ is the velocity operator and $\vert \Psi_{in} \rangle$
and $\vert \Psi_{fin} \rangle$ are, respectively, the states of a
particle before and after a velocity measurement.  We discuss the
consistency of weak values and show that superluminal weak speed is
consistent with relativistic causality.
\end{abstract}

\newpage
\section{Introduction}
     In quantum mechanics, it is axiomatic that the only allowed values
of an observable are its eigenvalues.  With these allowed values come,
in turn, allowed interpretations.  For example, a quantum particle can
tunnel through a potential energy barrier greater than its total energy.
Can it have negative kinetic energy?  The axiomatic answer is ``No! The
eigenvalues of kinetic energy are all positive!"  This answer does not
allow us an intuitive interpretation of quantum tunneling as a negative
kinetic energy phenomen.  But we can go beyond the axiomatic answer to
define the {\it weak} value $\langle A\rangle_w$ of an observable $A$
on a system \cite{spin,weak}:
\begin{equation}
\langle A\rangle_w ={{\langle \Psi_{fin} \vert A \vert \Psi_{in}
\rangle}\over {\langle \Psi_{fin} \vert \Psi_{in} \rangle}}~~~~.
\end{equation}
Here $\vert \Psi_{in} \rangle$ and $\vert \Psi_{fin}\rangle$ are,
respectively, the states of the system before and after a measurement
of $A$.  (Just as we can preselect $\vert \Psi_{in} \rangle$, we can
postselect $\vert \Psi_{fin} \rangle$; thus we measure $A$ on a pre-
and postselected ensemble.)  Weak values are measurable.  If the
measurement interaction is weak enough \cite{spin,weak}, measurements
on a pre- and postselected ensemble yield the weak value $\langle A
\rangle_w$; and $\langle A \rangle_w$ need {\it not} be an eigenvalue.
Indeed, it need not be any classically allowed value.  The weak kinetic
energy of a tunnelling particle is {\it negative} \cite{error}.  Weak
values allow many new interpretations, in addition to negative kinetic
energy.  Here we show that the weak speed of a particle can exceed
the speed of light, and we discuss the consistency of weak values.

     We will begin by showing how the weak speed of a charged particle
can exceed the speed of light {\it in vacuo}.  Such behavior seems
completely inconsistent with the laws of physics.  But we then compute
the electromagnetic field of the particle and find that it corresponds
to Cherenkov radiation:  like any charged particle moving faster
than light through a medium, a superluminal particle emits Cherenkov
radiation.  Finally, we prove that superluminal weak speed does not
contradict relativistic causality.  Weak speed illustrates the general
principle that all values measured on a pre- and postselected ensemble
are consistent.

\section{Quantum walk}

     Consider a particle constrained to move along the $z$-axis.
As a model Hamiltonian for our particle, we take $H= p_z v_z$, where
$p_z= -i\hbar \partial /\partial z$ and $v_z$ acts on an internal
Hilbert space of the particle:
\begin{equation}
v_z = {c\over{N}} \sum_{i=1}^N \sigma_z^{(i)}~~~~.
\end{equation}
The Pauli matrices operate on the internal Hilbert space.  (They do
not represent spin---the particle has no electric or magnetic dipole
moment.)  The eigenvalues of $v_z$ are $-c,-c+2c/N, \dots,c-2c/N,c$,
where $c$ is the speed of light.  The particle moves with velocity
$v_z$ in the $z$-direction,
\begin{equation}
{\dot x} = [x,H]/i\hbar = 0~~~,~~~~
{\dot y} = [y,H]/i\hbar = 0~~~,~~~~
{\dot z} = [z,H]/i\hbar = v_z~~~,
\end{equation}
hence the change in position $z$ is a measure of $v_z$.

     If the only allowed values of $v_z$ are its eigenvalues, the
speed of the particle cannot exceed the speed of light.  But consider
the following weak measurement of $v_z$.  We preselect the particle in
an initial state $\vert\Psi_{in} \rangle\Phi ({\bf x} ,0)$, where $\Phi
({\bf x} ,0)$ represents a particle approximately localized at ${\bf x}
=(x,y,z)=0$,
\begin{equation}
\Phi ({\bf x} ,0) =(\epsilon^2 \pi )^{-3/4} e^{-{\bf x}^2 /2\epsilon^2}~~~,
\label{phi}
\end{equation}
and postselect a final state $\vert\Psi_{fin}\rangle$.  For
$\vert\Psi_{in}\rangle$ and $\vert\Psi_{fin}\rangle$ we choose
\begin{eqnarray}
\vert\Psi_{in}\rangle &=&2^{-N/2} \otimes_{i=1}^N \left( \vert
\uparrow_i \rangle + \vert \downarrow_i \rangle  \right)~~~,\nonumber\\
\vert\Psi_{fin}\rangle &=&\otimes_{i=1}^N \left( \alpha_\uparrow \vert
\uparrow_i \rangle + \alpha_\downarrow \vert \downarrow_i \rangle  \right)~~~,
\label{psi}
\end{eqnarray}
with $\alpha_\uparrow$ and $\alpha_\downarrow$ real and
$\alpha_\uparrow^2 +\alpha_\downarrow^2 =1$.  Our chances of
postselecting the state $\vert \Psi_{fin}\rangle$ may be very small,
but if we repeat the experiment again and again, eventually we will
postselect $\vert \Psi_{fin} \rangle$.  Thus $\Phi ({\bf x},t)$ is
\begin{equation}
\Phi ({\bf x}, t)= \langle \Psi_{fin} \vert e^{-ip_z v_z
t/\hbar} \vert \Psi_{in}  \rangle \Phi ({\bf x} ,0) ~~~,\label{exp}
\end{equation}
up to normalization.  For short enough
times $t$, we can expand the exponent:
\begin{eqnarray}
\Phi ({\bf x}, t)&\approx &
\langle \Psi_{fin} \vert 1 -ip_z v_z t/\hbar \vert
\Psi_{in}  \rangle \Phi ({\bf x} ,0) \nonumber\\
&=&
\langle \Psi_{fin} \vert 1 -ip_z \langle v_z \rangle_w t/\hbar \vert
\Psi_{in}  \rangle \Phi ({\bf x} ,0) \nonumber\\
&\approx &
\langle \Psi_{fin} \vert e^{-ip_z \langle v_z\rangle_w t/\hbar} \vert
\Psi_{in}  \rangle \Phi ({\bf x} ,0) \nonumber\\
&=&
\langle \Psi_{fin} \vert\Psi_{in}  \rangle \Phi (x,y,z-\langle v_z
\rangle_w t,0) ~~~~. \label{taylor}
\end{eqnarray}
Thus at time $t$ the particle is displaced by $\langle v_z \rangle_w t$
along the $z$-axis.  Note that the weak value of $v_z$,
\begin{equation}
\langle v_z\rangle_w = {{\langle \Psi_{fin} \vert v_z\vert \Psi_{in}
\rangle }\over{\langle \Psi_{fin} \vert \Psi_{in} \rangle}} =
{{\alpha_\uparrow -\alpha_\downarrow}\over
{\alpha_\uparrow +\alpha_\downarrow}} c~~~, \label{wvalue}
\end{equation}
exceeds $c$ in magnitude if $\alpha_\uparrow \alpha_\downarrow$ is
negative.  Thus the weak speed of the particle could be superluminal.

     This result is so surprising as to merit a second derivation.
We can rewrite Eq.\ (\ref{exp}) by evaluating the exponent exactly:
\begin{eqnarray}
\Phi ({\bf x}, t)&=& 2^{-N/2}  \left( \alpha_\uparrow
e^{-ip_z ct/N\hbar}+ \alpha_\downarrow e^{ip_z ct/N\hbar}\right)^N \Phi
({\bf x},0) \nonumber\\
&=&2^{-N/2}  \sum_{n=0}^N \alpha_\uparrow^n
\alpha_\downarrow^{N-n} {{N!}\over{n!(N-n)!}} e^{-i(2n-N)p_z ct/N\hbar}
\Phi ({\bf x} ,0)~~~~.\label{bi}
\end{eqnarray}
Eq.\ (\ref{bi}) represents a superposition of many displacements of the
particle.  Applying the binomial theorem, we find that $\Phi ({\bf x},t)$
is a superposition of $\Phi ({\bf x},0)$ displaced along the $z$-axis by
at most $ct$ in either direction.  So how can Eq.\ (\ref{exp}) represent
a particle displaced by $\langle v_z\rangle_w t$ if $\langle v_z\rangle_w
t$ is out of this range?  Here is the surprise.  Apparently the displaced
states interfere, {\it constructively} for $z \approx \langle v_z
\rangle_w t$ and {\it destructively} for other values of $z$.  Indeed,
we can verify this interference.  Since
\begin{eqnarray}
\alpha_\uparrow e^{-ip_z ct /N\hbar} + \alpha_\downarrow e^{ip_zct /N\hbar}
&\approx &
\alpha_\uparrow (1-ip_zct /N\hbar) + \alpha_\downarrow (1 +ip_zct
/N\hbar)\nonumber\\
&=&
(\alpha_\uparrow +\alpha_\downarrow) -(\alpha_\uparrow -\alpha_\downarrow)
ip_zct /N\hbar \nonumber\\
&=&(\alpha_\uparrow +\alpha_\downarrow) (1 -i p_z \langle v_z \rangle_w
 t /N\hbar )~~~~\label{e1}
\end{eqnarray}
and
\begin{equation}
\lim_{N\rightarrow\infty} ( 1 - ip_z \langle v_z\rangle_w t /N\hbar )^N
=e^{-ip_z \langle v_z\rangle_w t /\hbar}~~~,\label{e2}
\end{equation}
we find that, for large enough $N$, Eq.\ (\ref{bi}) does indeed imply
Eq.\ (\ref{taylor}).

     Mathematically, Eq.\ (\ref{bi}) does not look like Eq.\
(\ref{taylor}).  Eq.\ (\ref{bi}) corresponds to a superposition of
waves $e^{-ip_z v_z t /\hbar}$ where $v_z= -c, -c+2c/N, \dots ,c-2c/N,c$.
If $e^{-i p_z \langle v_z\rangle_w t/\hbar}$ is not one of these waves,
how can we obtain it by superposing them?  Physically, Eq.\ (\ref{bi})
is analogous to a random walk.  We can generate a random walk in one
dimension by tossing a coin.  In Eq.\ (\ref{bi}), we toss a quantum
coin---a spin---to generate a quantum random walk \cite{qrandom}.  If
the coefficients $\alpha_\uparrow$ and $\alpha_\downarrow$ in Eq.\
(\ref{bi}) were probabilities, the expansion of Eq.\ (\ref{bi}) would
generate a classical random walk; each term in the expansion would
represent a possible random walk, with coefficient equal to its
probability.  A classical random walk of $N$ steps yields a typical
displacement of $\sqrt{N}$ steps, and never more than $N$.  But the
coefficients $\alpha_\uparrow$ and $\alpha_\downarrow$ are probability
amplitudes; the quantum random walk superposes all possible classical
random walks and yields arbitrary displacements.

\section{Consistency of weak values}

      We have derived Eq.\ (\ref{taylor}) two ways, but we have not
explained how such a surprising result as superluminal speed could
coexist with relativistic causality (i.e. the constraint $-c \le v_z
\le c$ that applies to $v_z$ and its eigenvalues.)  The explanation is
that superluminal speed depends on apparent ``errors" of measurement.
A hint of this dependence appears already in Eq.\ (\ref{phi}), where
we define the initial wave function $\Phi ({\bf x},0)$ of the particle
to be a gaussian with an uncertainty in position of about $\epsilon$.
If $\epsilon$ vanished, $\Phi ({\bf x},0)$ would be a delta-function
of position and no superluminal behavior could emerge from Eq.\
(\ref{bi}); there would be no tails on the wave function that could
interfere constructively for $z \approx \langle v_z\rangle_w t$.
However, $\epsilon$ does not vanish, our initial and final measurements
are uncertain, and we can obtain, ``by error", a displacement
corresponding to superluminal speed.  Thus  the weak value emerges
only if it could be an error; yet the weak value does not {\it seem}
to be an error.  On the contrary, whenever our pre- and postselections
(which are independent of $\Phi ({\bf x},0)$) yield the weak value
$\langle v_z \rangle_w$, measured values of the displacement of the
particle over a time $t$ cluster about $\langle v_z\rangle_w t$.

     We can quantify the dependence of weak speed on measurement
error as follows.  Eqs.\ (\ref{taylor}) and (\ref{bi}) agree in the
limit $N \rightarrow \infty$, but let us take into account the fact
that $N$ is finite.  To do so, we define a function $f(1/N) = (1+
s/N)^{N}$ with $s$ constant, and expand $f(1/N)$ in a Taylor series
expansion around $f(0)$:
\begin{equation}
f(1/N) = f(0) + {{f^\prime (0)}/ N} + {{f^{\prime\prime} (0)} /
{2N^2}} +\dots ~~~,
\end{equation}
where $f(0) =\lim_{N\rightarrow\infty} f(1/N)$, etc.  We obtain
\begin{equation}
\left( 1+{s \over N} \right)^N
 = e^s \left( 1-{{s^2}\over{2N}} +{{3s^4 +8s^3}\over
{24N^2}} +\dots \right)~~~~.
\end{equation}
Hence Eqs.\ (\ref{bi}) and (\ref{e1}) imply
\begin{equation}
\Phi ({\bf x}, t) =e^{-ip_z \langle v_z\rangle_w t/\hbar}
\left[ 1 + {{p_z^2 \langle v_z\rangle^2_w  t^2}\over{2N\hbar^2}} +{\cal{O}}
\left( {1 \over {N^2}} \right) \right] \Phi ({\bf x},0)~~~,
\end{equation}
up to normalization.  The exponential factor outside the brackets
displaces $\Phi ({\bf x},0)$ by $\langle v_z\rangle_w t$ but terms
of order $1/N$ can change the shape of $\Phi ({\bf x},0)$.  To make
the change negligible, we require
\begin{equation}
1\gg \langle v_z\rangle_w^2 t^2/N\epsilon^2~~~~.\label{cond}
\end{equation}
Eq.\ (\ref{cond}) relates $N$ to the width $\epsilon$ of $\Phi ({\bf x},
0)$:  to decrease $\epsilon$, we increase $N$.  As long as
Eq.\ (\ref{cond}) holds,  the particle will move with weak speed $\langle
v_z \rangle_w$ over a time $t$.

     Eq.\ (\ref{cond}) is crucial to the consistency of weak speed.
Does it seem that we get superluminal speed by playing a ``game
of errors" with the measuring device?  Perhaps; but it is a remarkably
consistent game:  whenever we preselect $\vert \Psi_{in}\rangle$ and
postselect $\vert\Psi_{fin}\rangle$ of Eq.\ (\ref{psi}), we get
superluminal speed (up to the uncertainty that characterizes the
measuring device).  For this consistency to hold, the probability of
postselecting $\vert\Psi_{fin}\rangle$ must be {\it smaller} than
the probability of getting the superluminal speed ``by error".
Otherwise, when we postselect $\vert\Psi_{fin} \rangle$, we would
most likely {\it not} get superluminal speed.

     Let's check:  on the one hand, a particle with wave function
$\Phi ({\bf x},0)$ may be found, by error, at $z= \langle v_z
\rangle_w t$ a time $t$ later. The probability of such an error is
proportional to $e^{-\langle v_z\rangle _w^2 t^2 /\epsilon^2}$,
which by Eq.\ (\ref{cond}) is much greater than $e^{-N}$.  On the
other hand, the probability of postselecting the state $\vert\Psi_
{fin}\rangle$ is approximately $(\alpha_\uparrow \alpha_\downarrow
+1/2)^N$.  If we compare the two probabilities and recall that
$\alpha_\uparrow\alpha_\downarrow$ is negative for $\langle v_z
\rangle_w >c$, we find that the probability of an error dominates
the probability of postselecting $\vert \Psi_{fin}\rangle$.

     Then why all the fuss about postselection?  If we measure $v_z$
and obtain the value $v_z >c$, what does it matter whether or not
we postselect?  The answer is that {\it only if we postselect are
measured values consistent}.  An example may help clarify this
answer.  Suppose we measure the displacement of the particle at time
$t$ with a weak measurement interaction.  If we don't postselect,
the most likely displacement at time $t$ is $ct/\sqrt{N}$, because
the expectation value of $v_z$ in the state $\vert \Psi_{in} \rangle$
is a random walk of $N$ steps of size $c/N$.  Yet there is a small
chance of obtaining a displacement $ct$.  Such a value might be an
error and, indeed, if we remeasure $z$, there is again only a
small chance of measuring such a large displacement.  Since each
measurement hardly disturbs the particle \cite{para}, the probability
that the next measurement yields a displacement $ct$ remains small.
Thus, without postselection, there is no consistency in measurement
errors.   Unless and until we postselect, they are just errors. With
postselection, however, measurement ``errors" yield a consistent
pattern.  Repeated weak measurements on an ensemble of particles
preselected in the state $\vert \Psi_{in}  \rangle$ and postselected
in the state $\vert \Psi_{fin}\rangle$ yield ``errors" consistent with
the superluminal weak value.

     Eqs.\ (\ref{taylor}) and (\ref{bi}) show that the weak speed of
a particle can consistently exceed $c$.  We now give the particle a
charge $q$ and show that its electromagnetic field, too, is consistent
with superluminal weak speed.

\section{Cherenkov radiation}

     What is the electromagnetic field of the particle?  Let us treat
the scalar potential; the treatment of the vector potential is similar.
To begin with, suppose that $v_z$ is well defined, {\it i.e.} that $v_z$
equals one of its eigenvalues.  Let $V({\bf x}^\prime ,t ;v_z)$ denote
the scalar potential at ${\bf x}^\prime ,t$ of a particle of charge $q$
moving along the $z$-axis with $z=v_z t$.  The simplest way to obtain
$V({\bf x}^\prime ,t ;v_z)$ is via a Lorentz boost, by $v_z$ in the
$z$-direction, of the Coulomb potential $V({\bf x}^\prime ,t;0)$.  We
obtain
\begin{equation}
V({\bf x}^\prime ,t ;v_z)
= q\left\{ [(x^\prime)^2 +(y^\prime)^2](1-v_z^2 /c^2 )
+(z^\prime -v_z t )^2 \right\}^{-1/2}~~~~.\label{vv}
\end{equation}

     So far, $V({\bf x}^\prime ,t ;v_z)$ represents the classical
potential of a point charge moving along the axis with $z=v_z t$.
But we want to treat the field as quantum mechanical.  We could do so
with quantum field operators, but the treatment would be unnecessarily
complicated.  Instead, let us write down an effective two-particle
interaction between the moving charge and a test particle.  Namely,
to $H=p_z v_z$, the Hamiltonian of the moving charge, we add the
Hamiltonian $H^\prime$ of a (nonrelativistic) test particle:
\begin{equation}
H^\prime = {1\over {2m}}\left( {\bf{p^\prime}} - q^\prime {\bf A}
\right)^2 + q^\prime V~~~~.\label{hprime}
\end{equation}
In $H^\prime$, the test particle has charge $q^\prime$, and the scalar
potential is
\begin{equation}
V({\bf x}^\prime )=
q\left\{  \left[ (x^\prime -x)^2
+(y^\prime -y)^2 \right] (1-v_z^2/c^2)+(z^\prime -z)^2 \right\}^{-1/2}~~~~.
\label{vee}
\end{equation}
The vector potential has only one nonzero component, namely $A_z$, which
is \cite{vec}
\begin{equation}
A_z ({\bf x}^\prime ) = {{qv_z}\over c}
\left\{  \left[ (x^\prime -x)^2
+(y^\prime -y)^2 \right] (1-v_z^2/c^2)+(z^\prime -z)^2 \right\}^{-1/2}~~~~.
\label{a}
\end{equation}
Note that if we substitute $(0,0, v_z t)$ for $(x,y,z)$, then $V({\bf
x}^\prime )$ equals $V({\bf x^\prime}, t ;v_z)$ as defined above in Eq.\
(\ref{vv}) and $A_z ({\bf x}^\prime )$ equals $(v_z/c) V({\bf x^\prime},
t ;v_z)$. The equations of  motion flowing from $H+H^\prime$ yield
$(x,y,z) =(0,0,v_z t)$ together with the correct motion of the test
particle due to the electromagnetic field of the moving charge. (The
equation of motion for the momentum ${\bf p}$ of the moving charge is
unphysical, but it has no measurable  consequences.)  Now we treat $V$
and $A_z$ as quantum operators and calculate their effect on the test
particle.  We will see that if the moving charge has weak speed
$\langle v_z \rangle_w$ then $\langle v_z \rangle_w$ replaces $v_z$
in Eqs.\ (\ref{vee}-\ref{a}).

     Namely, suppose we preselect the moving charge in the state $\vert
\Psi_{in} \rangle\Phi ({\bf x} ,0)$ and, after a time $T$, postselect
the state $\vert\Psi_{fin} \rangle$.  (See Eqs.\  (\ref{phi}-\ref{psi}).)
We also prepare the test particle in a localized state $\Omega ({\bf
x}^\prime ,0)$, where $\Omega ({\bf x}^\prime ,0)$ is analytic in ${\bf
x}^\prime$.  For simplicity, and because we want the test particle to
measure the instantaneous values of $A_z$ and $V$ at the end of this
evolution (and not their average values during or after the evolution),
we ``turn on" $H^\prime$ instantaneously at time $T$, i.e. we multiply
$H^\prime$ by $\delta (t-T)$.  The state of the moving charge and the
test particle after the postselection is then
\begin{equation}
\Phi ({\bf x}, T) \Omega ({\bf x}^\prime ,T)
= \langle \Psi_{fin} \vert
e^{-i[({\bf p}^\prime -q^\prime {\bf A})^2 /2m +q^\prime V]/\hbar}
e^{-ip_z v_z T/\hbar} \vert \Psi_{in} \rangle \Phi ({\bf x} ,0)
\Omega ({\bf x}^\prime ,0)~~~~.
\label{t}
\end{equation}
The potentials $V$ and ${\bf A}$ in Eq.\ (\ref{t}) are defined by
Eqs.\ (\ref{vee}-\ref{a}).  But we now show that the weak speed
$\langle v_z \rangle_w$ replaces $v_z$ in Eqs.\ (\ref{vee}-\ref{t}).
Here we present a short proof, while Appendix B contains a long
rigorous proof.

     Let us focus on the right-hand side of Eq.\ (\ref{t}) and note
that we can expand the first exponential,
\begin{equation}
e^{-i[({\bf p}^\prime -q^\prime {\bf A})^2 /2m +q^\prime V]/\hbar}~~~,
\end{equation}
as a power series in $v_z$.  Thus, the right-hand side of Eq.\
(\ref{t}) is a sum of terms of the form
\begin{equation}
\langle \Psi_{fin} \vert v_z^n e^{-ip_z v_z T/\hbar} \vert \Psi_{in}
\rangle
\end{equation}
multiplied on either side by functions that do not depend on $v_z$.
But we have, for any $n$ and in the limit $N\rightarrow\infty$,
\begin{eqnarray}
\langle \Psi_{fin} \vert v_z^n
e^{-ip_z v_z T /\hbar}
\vert \Psi_{in}
\rangle &=& \left( {{i\hbar} \over T}{\partial \over{\partial p_z}}
\right)^n
\langle \Psi_{fin} \vert
e^{-ip_z v_z T /\hbar} \vert \Psi_{in} \rangle
\nonumber\\
&=&
\langle \Psi_{fin} \vert \Psi_{in} \rangle
\left( {{i\hbar} \over T}{\partial \over{\partial p_z}}
\right)^n
e^{-ip_z \langle v_z \rangle_w T/ \hbar}
\nonumber\\
&=&
\langle \Psi_{fin} \vert \Psi_{in} \rangle
(\langle v_z \rangle_w )^n
e^{-ip_z \langle v_z \rangle_w T/\hbar} ~~~~.
\end{eqnarray}
(Compare Eqs.\ (\ref{exp}-\ref{e2}).)  So we can simply replace $v_z$ by
$\langle v_z\rangle_w$ everywhere it appears in the series.  We drop
the factor $\langle \Psi_{fin}\vert\Psi_{in}\rangle$ (to normalize)
and obtain
\begin{equation}
\Phi ({\bf x} ,T)  \Omega ({\bf x}^\prime ,T)
=
e^{-i [({\bf p}^\prime -q^\prime {\bf A} )^2 /2m +q^\prime V]/\hbar}
\Phi (x, y, z-\langle v_z \rangle_w T, 0)
\Omega ({\bf x}^\prime ,0)~~~,\label{gen}
\end{equation}
where
\begin{equation}
A_z = (\langle v_z \rangle_w /c)V = (\langle v_z \rangle_w /c)
V({\bf x}^\prime -{\bf x}, 0;\langle v_z \rangle_w )~~~~.\label{pen}
\end{equation}
Since $V({\bf x}^\prime -{\bf x}, 0;\langle v_z \rangle_w )$ equals
$V({\bf x}^\prime )$ as defined in Eq.\ (\ref{vee}) with $\langle v_z
\rangle_w$ taking the place of $v_z$, the scalar and vector potentials
are exactly the potentials of a charge moving with weak speed $\langle
v_z\rangle_w$ (folded with the width of the localized state $\Phi$) and
have the corresponding effect on the test particle.  Now if $\langle
v_z \rangle_w$ exceeds the speed of light, $V$ and $A_z$ correspond
to Cherenkov radiation, the shock wave of a charged particle moving
faster than light through a medium.

     Cherenkov radiation is a striking illustration of the principle
that all weak values measured on a pre- and postselected ensemble
are consistent.  There is more consistency here than what we have
noted.  We have shown that a particle emits weak Cherenkov radiation
consistent with its superluminal weak speed.  But we need not limit
ourselves to the Hamiltonian $H^\prime$ in Eq.\ (\ref{hprime}).
Given any Hamiltonian $H^\prime (v_z)$ that is analytic in $v_z$, we
can write the time evolution operator $e^{-i\int H^\prime (v_z )dt/
\hbar}$ as a power series in $v_z$, and then, as before, replace $
v_z$ by $\langle v_z \rangle_w$.  And what holds for weak speed holds
for other weak values.

     With our effective two-particle interaction $H^\prime$, we have
neglected the radiation modes of the electromagnetic field, just as
we often neglect these radiation modes in treating the interaction
between two charged particles via the Coulomb potential.  When can
we consistently neglect the radiation modes?  A particle of charge
$q$ reveals its position through its electromagnetic field; each
mode of the electromagnetic field is, in effect, a measuring device.
What assures us that the superposition of localized states in Eq.\
(\ref{t}) lasts a time $T$, if each localized state has a distinct
electromagnetic field?  In other words, how can we postselect the
state $\vert \Psi_{fin} \rangle$ if the radiation modes can reduce
the superposition to a localized state corresponding to one
eigenvalue of $v_z$?

     The answer to this question depends on the magnitude of the
charge $q$.  If $q$ is large, $\Phi ({\bf x},t)$ will not long remain
a superposition of localized states.  Each state in the superposition
corresponds to the charge moving at a different point along the
$z$-axis, localized to within $\Delta z \approx \epsilon$.  We assume
this uncertainty conforms to Eq.\ (\ref{cond}).  But if $q$ is large
enough, the radiation modes will measure the location of the charge
and reduce the uncertainty $\Delta z$ to less than what Eq.\
(\ref{cond}) allows, thereby reducing the superposition in Eq.\
(\ref{t}).  Conversely if $q$ is small, vacuum fluctuations will
dominate, and the radiation modes will not reduce the uncertainty
$\Delta z$ to less than what Eq.\ (\ref{cond}) allows.

     We can sharpen this question by imagining an observer at a
distance $D$ from the moving charge, who may or may not measure
its electric field to determine its position (and thus its speed).
If there is a measurement, it reduces the superposition in Eq.\ (\ref{t})
to a single localized state; then we cannot postselect $\vert \Psi_{fin}
\rangle$ and there will be no Cherenkov radiation.  But if there is
no measurement, and we postselect $\vert \Psi_{fin} \rangle$,
there {\it will} be Cherenkov radiation.  Can this observer violate
causality? As long as $D \le cT$, there is no problem:  the observer
is close enough to the particle to causally affect the outcome (whether
or not it emits Cherenkov radiation).  But for $D >cT$, the observer
cannot causally affect the particle before it emits Cherenkov radiation!
We are left with an apparent violation of causality; how can the
radiation from the particle be consistent with later measurements?

     To answer the question, let us suppose the observer locates the
particle by measuring its electric field.  At a distance $D$ from the
particle, the electric field strength $E$ is $E=q/D^2$, thus $D=
\sqrt{q/E}$.  Then $\Delta D =(D^3 /2q)\Delta E$.  Inferring the
position $z$ of the particle from this measurement of $E$, we have
$\Delta z \approx (D^3/2q) \Delta E$.  The condition for a weak
measurement of $v_z$ is Eq.\ (\ref{cond}), with $\Delta z$ taking the
place of $\epsilon$; that is,
\begin{equation}
\sqrt{N} (D^3 /2q) \Delta E \approx \sqrt{N} (\Delta z) \gg
\langle v_z\rangle_w T
~~~~.\label{delta}
\end{equation}
Since we assume $D\ge cT$, Eq.\ (\ref{delta}) implies $\sqrt{N} D^2
\Delta E \gg 2q$.  Now vacuum fluctuations in a region of volume $D^3$,
over a time $D/c$, induce uncertainty in the electric field that is
roughly $\Delta E \approx \sqrt{\hbar c} /D^2$ in magnitude
\cite{sakurai}.  Thus
\begin{equation}
\hbar c > 4q^2 /N \label{q}
\end{equation}
is the condition for weak measurement and Cherenkov radiation.  If
$q$ satisfies Eq.\ (\ref{q}), then weak Cherenkov radiation is
consistent with causality.  Indeed, even a strong interaction with
the electromagnetic field can show Cherenkov radiation:  for any given
$q$, $N$ must satisfy Eq.\ (\ref{q}), and then measurements will show
superluminal weak speed and Cherenkov radiation.  For $q\approx e$,
$N$ is approximately the inverse fine-structure constant; for larger
$q$, $N$ must be larger, as well.

     Thus Cherenkov radiation does not, by itself, imply superluminal
weak speed; we must still postselect $\vert\Psi_{fin} \rangle$.  Given
the condition $\hbar c > 4q^2 /N$, postselection of $\vert\Psi_{fin}
\rangle$  implies Cherenkov radiation, but the reverse does not hold:
Cherenkov radiation does not imply postselection of $\vert\Psi_{fin}
\rangle$.  Without postselection, Cherenkov radiation may be an error,
a fluctuation of the vacuum.

     In this example, we preselect $\vert \Psi_{in} \rangle$ and
postselect $\vert \Psi_{fin}\rangle$ to get superluminal weak speed.
In Eq.\ (\ref{psi}), which defines these states, all the coefficients
are real, and therefore the weak speed is real.  For other pre- and
postselections, however, the weak speed could be complex.  Complex
weak values can induce nonunitary time evolution.  An example we
will present elsewhere, of an imaginary weak dipole moment, shows a
remarkable interplay between imaginary weak values and entanglement.
Here, however, we discuss only real weak values.

\section{Relativistic causality}

     Weak measurements---measurements that yield weak values---are
internally consistent because they obey two rules.  On the one hand,
they are weak, hence they hardly disturb the measured system.  On the
other hand, they are inaccurate and can yield, ``by error", weak values.
These two rules are intimately related.  In our example,  the change
in the initial wave function $\Phi ({\bf x},0)$ is proportional to
$p_z$.  Thus, for the measurement to be weak, $p_z$ must be bounded.
But if $p_z$ is bounded, then the wave function is analytic \cite{error}
in $z$.  And since $\Phi ({\bf x} ,0)$ is analytic in $z$, the
probability density does not vanish for any interval in $z$.  Thus
we can localize the particle, ``by error", in a region it could not
have reached without superluminal speed.  What if we were to try to
eliminate the possibility of error, either by choosing the initial
wave function to be a Dirac delta function, or by otherwise imposing
a sharp cutoff on the initial wave function?  In either case, the
initial wave function would not be an analytic function.  But then
the expansion of Eqs.\ (\ref{taylor}) and (\ref{e1}) in powers of
$p_z$ would not be valid.  The exponential of $-ip_zv_z t/\hbar$ in
Eq.\ (\ref{exp}) is a unitary operator that translates $\Phi ({\bf x}
,0)$ to $\Phi (x,y, z-v_z t,0)$.  This unitary operator acts on any
wave function with a Fourier transform.  But the Taylor series
expansion of this unitary operator applied to $\Phi ({\bf x} ,0)$,
\begin{equation}
\sum_{m=0}^{\infty} {{(-ip_z v_zt/\hbar )^m}\over{m!}} \Phi ({\bf x} ,0)~~~,
\end{equation}
equals the Taylor series expansion of $\Phi (x,y,z-v_z t,0)$ around
$\Phi ({\bf x} ,0)$ only if $\Phi ({\bf x} ,0)$ is an analytic function.
Thus the weak value $\langle v_z \rangle_w$ emerges in this experiment
only if the initial wave function $\Phi ({\bf x} ,0)$ is analytic.

     Once we understand the role of analyticity in the emergence of
$\langle v_z\rangle _w$, we can answer another question:  How can
$\langle v_z\rangle_w > c$ be consistent with relativistic causality?
We have seen that the particle moves with velocity $\langle v_z
\rangle_w$ only if $\Phi ({\bf x} ,0)$ is analytic.  But if $\Phi ({\bf
x} ,0)$ is analytic, then its value and the value of its derivatives at
any one point determine its value at all points.  Hence $\Phi ({\bf x}
,t) =\Phi (x,y,z -\langle v_z\rangle_w t ,0)$ does not transmit any
message, because it is the same message for all ${\bf x}$ and $t$.
Since $\Phi ({\bf x},t)$ does not transmit any message, it does not,
in particular, transmit a superluminal message, and there is no
violation of relativistic causality.

     Thus superluminal weak speed is consistent with relativistic
causality and with other measurements.  There are two distinct ways
in which weak measurements can be consistent.  On the one hand, if a
weak measurement of $v_z$ on a pre- and postselected ensemble yields
$\langle v_z \rangle_w >c$, any weak measurement of the electromagnetic
field {\it on the same pre- and postselected ensemble} will show
Cherenkov radiation.  That is, weak measurements are consistent as
long as they apply to the same pre- and postselected ensemble.  On
the other hand, if measurements do not apply to the same pre- and
postselected ensemble, they are consistent even if they yield different
measured values.  For example, we can follow a weak measurement of
$v_z$ with either a postselection or a precise measurement of $v_z$.
If we postselect the state $\vert \Psi_{fin}\rangle$, we interpret the
result of the weak measurement as the weak value $\langle v_z \rangle_w$;
if we precisely (re)measure $v_z$, we may interpret the result of the
weak measurement as an error.  But these two interpretations of a
measured value are  consistent, for they apply to different
ensembles---the former to a  pre- and postselected ensemble and the
latter to a preselected ensemble.  Thus, how we interpret a measured
value depends on what we choose to measure next.  Here we have
considered weak measurements on a single pre- and postselected
ensemble. Together, these measurements yield a consistent picture
of a charge moving in the vacuum at superluminal speed and emitting
Cherenkov radiation.

\acknowledgments
We thank a referee for comments that helped us write more clearly.

\appendix

\section{}

We will prove \cite{jackson} the following representation for
$V({\bf x}^\prime ,t;v_z)$:
\begin{equation}
V({\bf x}^\prime ,t ;v_z) =q\int_{-\infty}^{\infty} d\tau
{{\delta (t -\tau -\vert {\bf x}^\prime -{\bf x}\vert /c)}\over
{\vert{\bf x}^\prime -{\bf x}\vert}}
~~~~.\label{v}
\end{equation}
Here $\vert {\bf x}^\prime -{\bf x} \vert = [(x^\prime )^2 +(y^\prime )^2
+(z^\prime -v_z \tau )^2 ]^{1/2}$.  We evaluate the $\delta$-function at
its zeros according to the rule
\begin{equation}
\delta (g(\tau )) = \sum_i {{\delta (\tau -\tau_i )}\over{\vert
dg(\tau )/d\tau \vert}}~~~,
\end{equation}
where $\tau_i$ satisfies $g(\tau_i ) =0$ and here
\begin{equation}
g(\tau ) = t-\tau -[(x^\prime )^2 +(y^\prime )^2 +(z^\prime -v_z \tau
)^2
]^{1/2} /c~~~~.
\end{equation}
To obtain the zeros, we solve the quadratic equation
\begin{equation}
c^2 (t -\tau )^2 = (x^\prime)^2 + (y^\prime )^2 +(z^\prime -v_z\tau )^2
\end{equation}
and require $t \ge \tau$.  There is one zero for $\vert v_z\vert
<c$,
\begin{equation}
c\tau ={{ct -v_z z^\prime /c +\left\{ [(x^\prime)^2 +(y^\prime)^2]
(1-v_z^2 /c^2)
+(z^\prime -v_z t )^2 \right\}^{1/2}}\over{1-v_z^2 /c^2}}~~~,
\end{equation}
and the integral yields
\begin{equation}
V({\bf x}^\prime ,t ;v_z)
= q\left\{ [(x^\prime)^2 +(y^\prime)^2](1-v_z^2 /c^2 )
+(z^\prime -v_z t )^2 \right\}^{-1/2}~~~,
\end{equation}
as before.  This representation of $V({\bf x}^\prime ,t  ;v_z)$ will be
very useful in Appendix B.

\section{}

We will show that the weak speed $\langle v_z \rangle_w$ replaces $v_z$
in $V$ and ${\bf A}$ in Eq.\ (\ref{t}).  We first show it in the limit
$m\rightarrow\infty$, i.e. we first consider only the scalar potential
$V$.  Then we generalize to finite $m$ and consider ${\bf A}$ too.

     Let us focus on the term in angle brackets in Eq.\ (\ref{t}) and
begin by noting that $V ({\bf x}^\prime )$ as defined in Eq.\ (\ref{vee})
can also be written $V( {\bf x}^\prime -{\bf x},0; v_z )$ as defined
in Eq.\ (\ref{vv}).  Hence (in the limit $m\rightarrow\infty$) we can
write the term in angle brackets as
\begin{eqnarray}
\langle \Psi_{fin} \vert
e^{-i q^\prime V/\hbar}
e^{-ip_z v_z T/\hbar} \vert \Psi_{in} \rangle
&=&
\langle \Psi_{fin} \vert
e^{-i q^\prime V({\bf x}^\prime -{\bf x}, 0;v_z )/\hbar}
e^{-ip_z v_z T/\hbar} \vert \Psi_{in} \rangle \nonumber \\
&=&
\langle \Psi_{fin} \vert
e^{-ip_z v_z T/\hbar}
e^{-i q^\prime V({\bf x}^\prime -{\bf x}, T;v_z )/\hbar}
 \vert \Psi_{in} \rangle ~~~~.
\end{eqnarray}
The trick is to take the dependence on $v_z$ out of $V({\bf x}^\prime
-{\bf x}, T;v_z )$ and put it in a more convenient place.  To this
end, we refer to the representation in Eq.\ (\ref{v}) and note that
all the dependence on $v_z$ is contained in the expression $\vert
{\bf x}^\prime -{\bf x} \vert$ which, for $V({\bf x}^\prime -{\bf x},
T;v_z )$, equals $[(x^\prime -x)^2 +(y^\prime -y)^2 +(z^\prime -z
-v_z \tau )^2 ]^{1/2}$.  It follows that the combination
\begin{equation}
M=e^{ip^\prime_z v_z \tau /\hbar}
e^{-i q^\prime V({\bf x}^\prime -{\bf x}, T; v_z )/\hbar}
e^{-ip^\prime_z v_z \tau /\hbar}
\end{equation}
is actually {\it independent} of $v_z$ and we can write the term in
angle brackets as
\begin{equation}
\langle \Psi_{fin} \vert
e^{-ip_z v_z T/\hbar}
e^{-ip^\prime_z v_z \tau /\hbar}
M
e^{ip^\prime_z v_z \tau /\hbar}
\vert \Psi_{in} \rangle ~~~,\label{tmp}
\end{equation}
where $M$ is independent of $v_z$.  We would like to move $M$ out of the
angle brackets.  Indeed we can do so, even though $M$ does not commute
with $e^{ip^\prime_z v_z \tau /\hbar}$.  The reason is that we can
always write $\Omega ({\bf x}^\prime ,0)$ as a sum of Fourier components.
For each Fourier component in the sum, we can move $M$ out of the
angle brackets, and later move it back in; hence we can do so for the
sum itself. Thus we can rewrite Eq.\ (\ref{tmp}) as
\begin{equation}
\langle \Psi_{fin} \vert
e^{-ip_z v_z T/\hbar}
e^{-i( p^\prime_z -{\bar p}^\prime_z )v_z \tau /\hbar}
\vert \Psi_{in} \rangle M
=
\langle \Psi_{fin} \vert \Psi_{in} \rangle
e^{-ip_z \langle v_z\rangle_w T/\hbar}
e^{-i(p^\prime_z -{\bar p}^\prime_z )\langle v_z\rangle_w \tau /\hbar}
M~~~,
\end{equation}
where ${\bar p}^\prime_z$ represents an eigenvalue of $p^\prime_z$ for
a given Fourier component.  (We have taken the limit $N\rightarrow
\infty$; compare Eqs.\ (\ref{exp}-\ref{e2}).)  Now we can pull $e^{i{\bar
p}^\prime_z \langle v_z\rangle_w \tau /\hbar}$ back to the right side
of $M$, turn ${\bar p}^\prime_z$ back into $p^\prime_z$, drop the factor
$\langle \Psi_{fin} \vert \Psi_{in} \rangle$ (to normalize), and rewrite
the term in angle brackets as
\begin{equation}
e^{-ip_z \langle v_z\rangle_w T/\hbar}
e^{-ip^\prime_z \langle v_z\rangle_w \tau /\hbar}
M
e^{ip^\prime_z \langle v_z\rangle_w \tau /\hbar}
=
e^{-ip_z \langle v_z\rangle_w T/\hbar}
e^{-i q^\prime V({\bf x}^\prime -{\bf x}, T;\langle v_z \rangle_w )
/\hbar}~~~~. \label{good}
\end{equation}
Applying Eq.\ (\ref{good}) to the combined state $\Phi ({\bf x} ,0)
\Omega ({\bf x}^\prime ,0)$ of the moving charge and the test particle,
we obtain at time $T$
\begin{eqnarray}
\Phi ({\bf x} ,T)  \Omega ({\bf x}^\prime ,T)
&=& e^{-ip_z \langle v_z\rangle_w T/\hbar}
e^{-i q^\prime V({\bf x}^\prime -{\bf x}, T;\langle v_z \rangle_w )
/\hbar}
\Phi ({\bf x} ,0)  \Omega ({\bf x}^\prime ,0) \nonumber \\
&=&
e^{-i q^\prime V({\bf x}^\prime -{\bf x}, 0;\langle v_z \rangle_w )
/\hbar}
\Phi (x, y, z-\langle v_z \rangle_w T, 0)
\Omega ({\bf x}^\prime ,0) ~~~~.\label{evo}
\end{eqnarray}
Eq.\ (\ref{evo}) corresponds to Eqs.\ (\ref{gen}-\ref{pen}) in the limit
$m\rightarrow\infty$.

     Now let $m$ be finite.  Since $A_z ({\bf x}^\prime )$ equals $(v_z
/c) V({\bf x}^\prime )$,  we can define a representation of $A_z ({\bf
x}^\prime, t ;v_z )$ to be $(v_z /c)$ times the representation of $V
({\bf x}^\prime , t ;v_z )$ in Eq.\ (\ref{v}).  But how do we deal with
this extra dependence on $v_z$ in $A_z ({\bf x}^\prime, t ;v_z )$?
We can expand the exponential term
\begin{equation}
e^{-i[({\bf p}^\prime -q^\prime {\bf A})^2 /2m +q^\prime V]/\hbar}
\end{equation}
in Eq.\ (\ref{t}) as a Taylor series.  If we then replace $A_z ({\bf
x}^\prime )= (v_z /c) V ({\bf x}^\prime -{\bf x} , 0 ;v_z )$ by its
representation, there will be powers of $v_z$ in the series.  But
we have, for any $n$ and in the limit $N\rightarrow\infty$,
\begin{eqnarray}
\langle \Psi_{fin} \vert v_z^n
e^{-i(p_z T + p^\prime_z \tau -{\bar p}^\prime_z \tau )v_z/\hbar}
\vert \Psi_{in}
\rangle &=& \left( {{i\hbar} \over T}{\partial \over{\partial p_z}}
\right)^n
\langle \Psi_{fin} \vert
e^{-i(p_z T + p^\prime_z \tau -{\bar p}^\prime_z \tau )v_z/\hbar}
\vert \Psi_{in} \rangle
\nonumber\\
&=&
\langle \Psi_{fin} \vert \Psi_{in} \rangle
\left( {{i\hbar} \over T}{\partial \over{\partial p_z}}
\right)^n
e^{-i(p_z T + p^\prime_z \tau -{\bar p}^\prime_z \tau )\langle
v_z \rangle_w/\hbar}
\nonumber\\
&=&
\langle \Psi_{fin} \vert \Psi_{in} \rangle
(\langle v_z \rangle_w )^n
e^{-i(p_z T + p^\prime_z \tau -{\bar p}^\prime_z \tau )\langle
v_z \rangle_w/\hbar} ~~~,\label{trick}
\end{eqnarray}
so we can replace $v_z$ by $\langle v_z\rangle_w$ everywhere it
appears in the series.  Then we obtain Eqs.\ (\ref{gen}-\ref{pen}) as
the generalization of Eq.\ (\ref{evo}).


\begin{references}
\bibitem{para}Suppose at time $t_0$ we measure $v_z t_0$ with precision
$ct_0 /\sqrt{N}$, e.g. the interaction Hamiltonian for the measurement
is $H_{int} =\delta (t-t_0) P_d v_z \sqrt{N}/c$, where $P_d$ is the
momentum conjugate to the position $Q_d$ of a pointer on the measuring
device.  Then since $v_z = (c/N) \sum \sigma_z^{(i)}$, the strength
of the interaction with each state $(\vert \uparrow_i \rangle +\vert
\downarrow_i\rangle )/\sqrt{2}$ is  proportional to $1/\sqrt{N}$; and
the probability that the measurement leaves this state unchanged
equals the expectation value $\langle \cos^2 (P_d /\sqrt{N} )\rangle$.
The probability that the measurement leaves the whole state $\vert
\Psi_{in} \rangle$ unchanged is $\langle \cos^{2N} (P_d/\sqrt{N} )
\rangle$ which, for large $N$, approaches $\langle e^{-P_d^2}\rangle$
and can be arbitrarily close to 1.
\end{references}
\end{document}